# What limits photosynthesis?
# Identifying the thermodynamic constraints of the terrestrial biosphere within the Earth system

Axel Kleidon

Max-Planck-Institute for Biogeochemistry

Hans-Knöll-Str. 10

D-07745 Jena, Germany

e-mail: akleidon@bgc-jena.mpg.de

**Abstract**

Photosynthesis converts sunlight into the chemical free energy that feeds the Earth's biosphere, yet at levels much lower than what thermodynamics would allow for. I propose here that photosynthesis is nevertheless thermodynamically limited, but this limit acts indirectly on the material exchange. I substantiate this proposition for the photosynthetic activity of terrestrial ecosystems, which are notably more productive than the marine biosphere. The material exchange for terrestrial photosynthesis involves water and carbon dioxide, which I evaluate using global observation-based datasets of radiation, photosynthesis, precipitation and evaporation. I first calculate the conversion efficiency of photosynthesis in terrestrial ecosystems and its climatological variation, with a median efficiency of 0.77% ($n$ = 13274). The rates tightly correlate with evaporation on land ($r^2$ = 0.87), which demonstrates the importance of the coupling of photosynthesis to material exchange. I then infer evaporation from the maximum material exchange between the surface and the atmosphere that is thermodynamically possible using datasets of solar radiation and precipitation. This inferred rate closely correlates with the observation-based land evaporation dataset ($r^2$ = 0.84). When this rate is converted back into photosynthetic activity, the resulting patterns correlate highly with the observation-based dataset ($r^2$ = 0.66). This supports the interpretation that it is not energy directly that limits terrestrial photosynthesis, but rather the material exchange that is driven by sunlight. This interpretation can explain the very low, observed conversion efficiency of photosynthesis in terrestrial ecosystems as well as its spatial variations. More generally, this implies that one needs to take the necessary material flows and exchanges associated with life into account to understand the thermodynamics of life. This, ultimately, requires a perspective that links the activity of the biosphere to the thermodynamic constraints of transport processes in the Earth system.

**Introduction**

Photosynthesis is the most dominant process by which chemical free energy is generated in the Earth's system (Kleidon 2016) and which sustains the Earth's biosphere. This chemical free energy, and the associated chemical disequilibrium, is reflected in the high concentration of oxygen in the Earth's atmosphere and the large amounts of reduced, organic carbon compounds elsewhere, such as the biomass associated with the biosphere, organic carbon stored in soils, and hydrocarbons contained in geologic reservoirs. This energy has substantially transformed the physical and chemical environment of the Earth, from covering tropical regions with lush rainforests to transforming an atmosphere to low greenhouse gas concentrations, particularly of carbon dioxide, and high levels of reactive oxygen. We may ask which factors ultimately constrain the level of photosynthetic activity? Are the constraints the kinetic reaction constants at the molecular scale, constraints to biological evolution, environmental factors, or the fundamental laws of thermodynamics?

What I want to propose here is that the answer likely lies in the combination of the latter two factors, that is, that the laws of thermodynamics limit photosynthetic activity, but that this limit acts through environmental factors rather than directly on the energy conversion process from solar radiation into the chemical free energy stored in carbohydrates. To illustrate this proposition, I focus on the photosynthetic activity on land, as terrestrial ecosystems at large scales are substantially more productive than the marine counterparts. This is reflected at the planetary scale in estimates of how much carbon is taken up in form of $CO_2$ by photosynthesis. While the marine biosphere takes up around $50 \times 10^{15}$ grams of carbon per year (Ciais et al., 2013) over about three quarters of the planetary surface that are covered by oceans, the terrestrial biosphere takes up more than twice as much ($123 \times 10^{15}$ grams of carbon per year, Ciais et al., 2013), yet over much less surface area. The highly productive rainforest ecosystems are thus a good reference point for understanding which factors constrain their high rates of photosynthesis and how these relate to thermodynamics. This focus on carbon uptake may not cover all parts of biotic activity, as it leaves out dynamics that are directly driven by the energy stored in form of ATP. The focus on land involves specific representations of the material exchange in form of carbon and water that are quite different in the marine counterpart, where water is not a limitation. I would nevertheless argue that the focus on carbohydrates as a means to evaluate the limits of photosynthesis is adequate because carbohydrates provide the fuel used for plant growth, reproduction, and to sustain further trophic levels in food chains. It thus describes the part of the chemical free energy generated by photosynthesis that sustains life and transforms the planetary environment. While the material exchange for terrestrial photosynthesis is specific to land, material exchange may still be generally limiting other forms of life, although this form of limitation may manifest itself differently.

The direct route by which thermodynamics could limit photosynthesis through energy conversion has been investigated for decades (Duysens 1962, Press 1976, Radmer and Kok 1977, Landsberg and Tonge 1980). These studies repeatedly showed that actual photosynthetic rates observed in the natural environment are substantially lower than what a thermodynamic limit would predict. The starting point for these evaluations is the very low entropy contained in solar radiation. Its low entropy is associated with the high radiative emission temperature of the Sun, and it is reflected in a radiative flux with relatively few, energy-rich photons with wavelengths centered in the visible wavelength range. Once absorbed at the temperatures of the Earth's surface and remitted to space, this energy is contained in a radiative flux of the same magnitude, but of many more, but less energetic photons with longer wavelengths in the infrared range. This difference in the entropy of radiative fluxes that are absorbed and emitted by the Earth system establishes the entropy exchange that allows for substantial dissipative activity on Earth (Boltzmann 1886, Ozawa et al. 2003, Kleidon and Lorenz 2005, Kleidon 2010).

Photosynthesis can, in principle, be highly effective in capturing the low-entropy solar photons and use these to generate chemical free energy in photochemical conversions before these photons get degraded to heat at ambient temperatures of Earth's environment. Photosynthesis uses a minimum of 8 moles of photons of 680 and 700 nm wavelengths to fix one mole of carbon dioxide (Hill and Rich, 1983). With each photon containing about 1.8 eV of energy, this yields at least an energy amount of 14.4 eV. With 1 eV = 1.6 x 10$^{-19}$ J, and the Avogadro number of $N_0$ = 6.022 x 10$^{23}$, this yields an energy requirement of 1387.5 kJ per mole of carbon (mol C) or 115.6 kJ gC$^{-1}$ (grams of Carbon). Carbohydrates contain about 470 kJ/mol C of energy, derived from 2.8 MJ/mol contained in glucose (Atkins and de Paola, 2010), which contains 6 carbon atoms, which combined yields a conversion efficiency of about 34% (= 470 kJ/1387.5 kJ per mol C). Hill and Rich (1983) found that this conversion efficiency is very close to the theoretical maximum efficiency of 36% and can be found in the light response curve of photosynthesis under low light conditions. However, only about 55% of solar radiation contains wavelengths that can be utilized by photosynthesis, reducing the maximum efficiency to about 18%. When looked at photosynthesis in actual ecosystems, Monteith (1972, 1977) found that these have a much lower efficiency, and convert less than 3% of the absorbed solar radiation into chemical free energy. This low efficiency of photosynthesis is well recognized, also in agricultural research as it is associated with low limits to agricultural yields (e.g., Zhu et al. 2010). Photosynthesis thus appears to operate well below the thermodynamic limit of converting sunlight into chemical free energy.

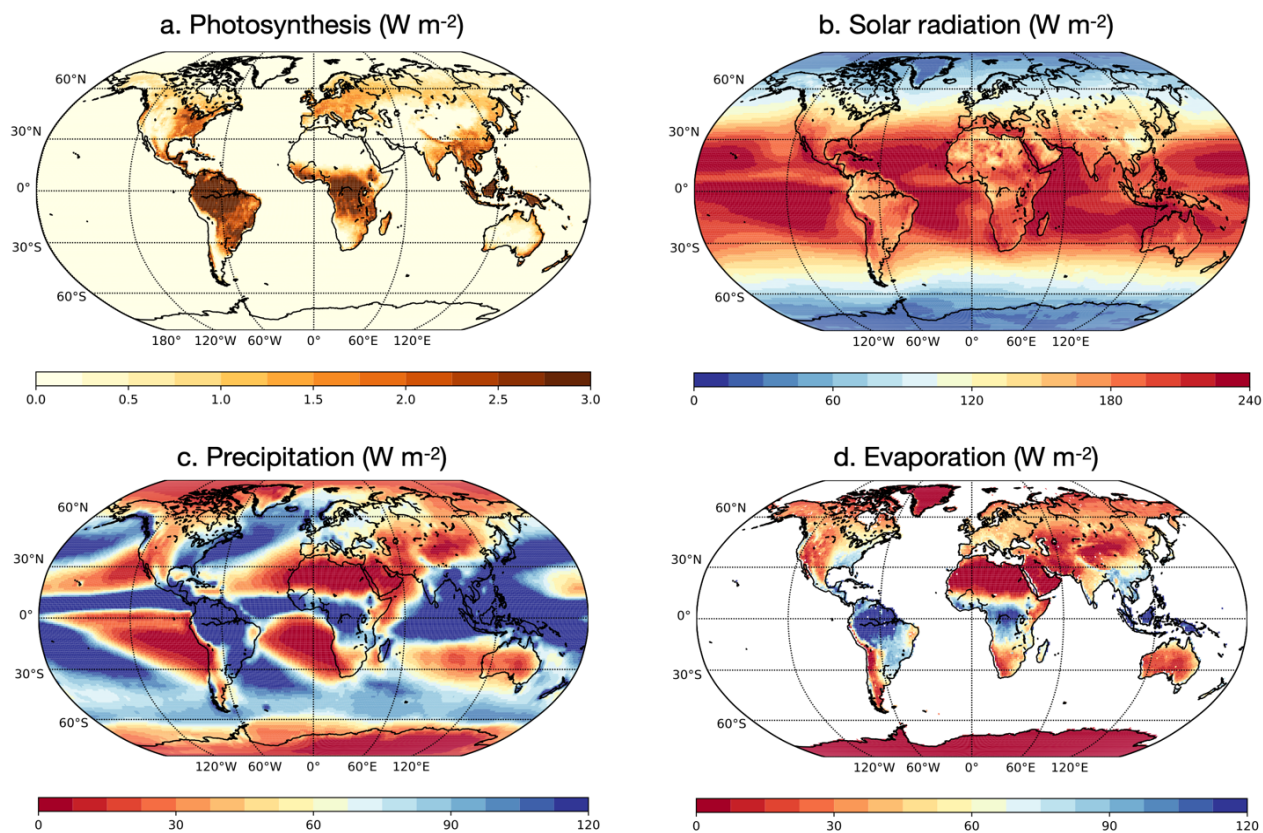

**Figure 1:** Photosynthetic activity of the terrestrial biosphere derived from satellite observations (a.) and its relation to (b.) absorbed solar radiation, (c.) precipitation, and (d.) evaporation on land. All maps show observational datasets (Table 1), with the units converted to W m$^{-2}$ and the details given in the Methods and Materials section.

Even though thermodynamics does not constrain the energy conversion by photosynthesis, the climatological variation of gross carbon uptake in the natural environment nevertheless shows systematic and predictable variations. These variations are shown in Figure 1, using an estimate of photosynthetic carbon uptake by terrestrial ecosystems that is based on satellite observations (Randerson et al 2017, Ott 2020). For comparison, datasets on absorbed solar radiation (Loeb et al. 2018, Kato et al. 2018), precipitation (Adler et al., 2016), and evaporation (Miralles et al, 2011; Martens et al, 2017) are also shown in Figure 1. I converted these datasets to energy units to make them comparable (see methods section below for detail), so that the rate of photosynthesis is given in terms of how much chemical free energy is generated, and the rates of precipitation and evaporation relate to how much latent heat is taken up or released during the phase conversions. When viewed in energy units, it shows that photosynthesis constitutes a comparatively small flux of energy compared to the other variables shown in Figure 1.

What Figure 1 also shows is that the geographic variations of photosynthesis clearly co-vary with environmental factors, specifically regarding the availability of water (or, precipitation) and light (or, solar radiation). These clear variations have led to the concepts of light- and water limited regimes of photosynthetic activity in natural ecosystems (Monteith 1972, also Potter et al. 1993, Churkina and Running 1998, Beer et al. 2010). Yet, if photosynthesis could, in principle, use much more of the energy contained in light, as suggested by the thermodynamic considerations from above, why do we refer to light-limited rates of photosynthesis in terrestrial ecosystems when water is not the limiting factor?

Here I want to describe an interpretation of the limits of photosynthetic activity on land that brings together thermodynamic constraints and the environmental context. At the center of this interpretation is the notion that in order to maintain any form of chemical reaction in the environment, including those that are central to life, it does not just need reactants, products, and energy. It also requires transport, or better, material exchange, that keeps supplying the reactants and keeps removing the products from the reaction. This, of course, also applies to the metabolic activity of living organisms. For me, this view is at the heart of the term "biogeochemistry" - the integration of life with the geochemical transformations and the geophysical exchange processes of the Earth system. The material exchange process in itself also results from energy conversions, as motion involves kinetic energy, and this energy needs to come from somewhere (i.e., it needs to have been converted from an energy source). These physical conversions may appear at first sight unconnected from the geochemical reaction or the metabolic activity of the biosphere. Yet, what I want to show here is that these connections play a central role.

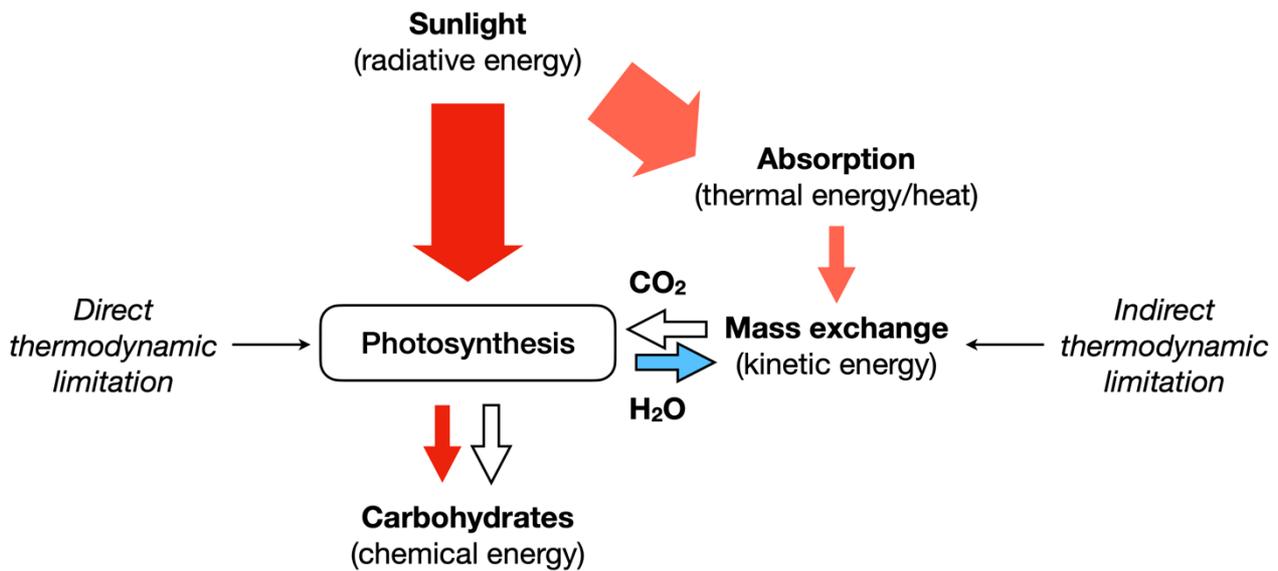

Figure 2: Schematic diagram illustrating the role of thermodynamics in limiting photosynthesis (left) directly through the photochemical conversion process from radiative to chemical free energy and (right) indirectly through material exchange of the reactants and products with the environment that is linked to kinetic energy derived from radiative heating by sunlight. This paper focuses on the indirect limitation shown as the chain of arrows on the right.

More specifically, when we deal with photosynthetic activity on land, we deal with the mass exchange of carbon dioxide and water between the plant canopies and the surrounding atmosphere. While plants take up atmospheric carbon dioxide, which diffuses into the leaves through their stomata, they inevitably lose water that diffuses in the opposite direction from the nearly saturated air space of the leaves' interior to the atmosphere. This exchange through the stomata is driven by the respective differences in concentrations across the canopy-atmosphere interface. It may seem that if the air was saturated, with a relative humidity of 100%, there is no loss of water as there is no difference to level out and plants could take up carbon dioxide without water loss. Yet, in natural environments, the relative humidity typically declines during the day as the surface is heated by absorption of solar radiation, warming the near-surface air. This makes the air drier and drives transpiration. On the other hand, the photosynthetic carbon uptake at the surface leaves a marked imprint in the atmosphere, lowering the concentrations of carbon dioxide in the convective boundary layer during the day (e.g., Wofsy et al., 1988). These environmental

processes – absorption of solar radiation, the generation of convective motion, transpiration and photosynthetic carbon uptake are thus tightly linked, although how these are linked can vary to some extent by variations in the meteorological conditions.

This linkage of carbon gain and water loss associated with the gas exchange of plant canopies results in a strong correlation, as can be seen by the close correspondence of the maps of photosynthesis and evaporation (compare Fig. 1a to 1d). It is captured by the well-established concept of water use efficiency, which describes how much water plants lose as they take up carbon. This strong coupling of metabolic activity and the mass exchange of different gases between organisms and their environment is well recognized and reflected in observations, not just for plants, but also for animals (Woods and Smith, 2010). Note also that this water loss by transpiration is substantially greater than the actual need for water in the photosynthetic reaction.

This is where thermodynamics and its limits comes back in. Evaporation from the land surface into the atmosphere requires water input by precipitation, energy input for the phase transition, but also the exchange of the moistened air at the surface with the drier air aloft. This exchange is generated mostly by the heating of the surface by absorbing sunlight, generating buoyant air that rises and takes the evaporated water from the surface into the higher atmosphere. Thermodynamics limits how much of this convective exchange can be generated, noting that this exchange requires kinetic energy, and this kinetic energy is generated similarly to a heat engine that operates between the heated surface and the cold atmosphere. The thermodynamic limit predicts how much of the absorbed radiation at the surface is being transported to the atmosphere in form of sensible and latent heat, and can therefore predict evaporation rates (Kleidon and Renner, 2013a, Kleidon et al. 2014, Conte et al. 2019) as well as its response to climate change (Kleidon and Renner, 2013b, Kleidon and Renner 2017). As water loss from the surface is so tightly linked to carbon uptake, as shown in Figure 1, what this suggests is that this form of thermodynamic limit also acts to indirectly constrain photosynthesis (Figure 2). What I propose here is that thermodynamics does not limit photosynthesis through the direct conversion of sunlight into chemical free energy, but indirectly through setting limits to the necessary material exchange between photosynthesizing plant canopies and the overlying atmosphere that allows them to simultaneously transpire water and take up carbon dioxide (also, Kleidon 2016). This thermodynamic limitation on the simultaneous exchange of water and carbon then results in the tight correlation between transpiration and carbon uptake, as shown in the maps in Figure 1 and evaluated in greater detail below.

In the following, I extend this line of interpretation about the limiting factors of photosynthetic activity on land using observational datasets. The methods section describes the thermodynamic limit on the material exchange between the land surface and the atmosphere that is shown in Figure 2 (right), summarizing earlier work (Kleidon and Renner 2013a,b; Kleidon et al. 2014; Kleidon and Renner 2017). This section also describes the satellite-derived datasets from Figure 1 that were used to quantify the thermodynamic efficiency, the tight linkage to evaporation, and the thermodynamic limit. The evaluation of the datasets and the thermodynamic limit is then presented in the results section. The discussion describes potential shortcomings, links the results to the proposed interpretation of the indirect thermodynamic limitation of photosynthetic activity on land, and describes the consistency of this interpretation with other studies. I close with a discussion on the broader implications of this interpretation of biotic activity from thermodynamics and exchange limitations, including the origin of life and its evolution and emphasizing the importance of an Earth system approach. I close with a brief summary and conclusions.

**Methods and Materials**

In this section, I first describe the background of the thermodynamic limit of maximum power, which predicts the extent to which heat and mass is exchanged vertically between the surface and the atmosphere. This limit results in an expression that predicts evaporation from the radiative forcing of solar radiation when sufficient water is accessible at the surface. It represents a thermodynamic limit for mass exchange that should also apply to carbon dioxide exchange for photosynthesis, so it forms the basis of the key argument shown in Figure 2 and for the following evaluations.

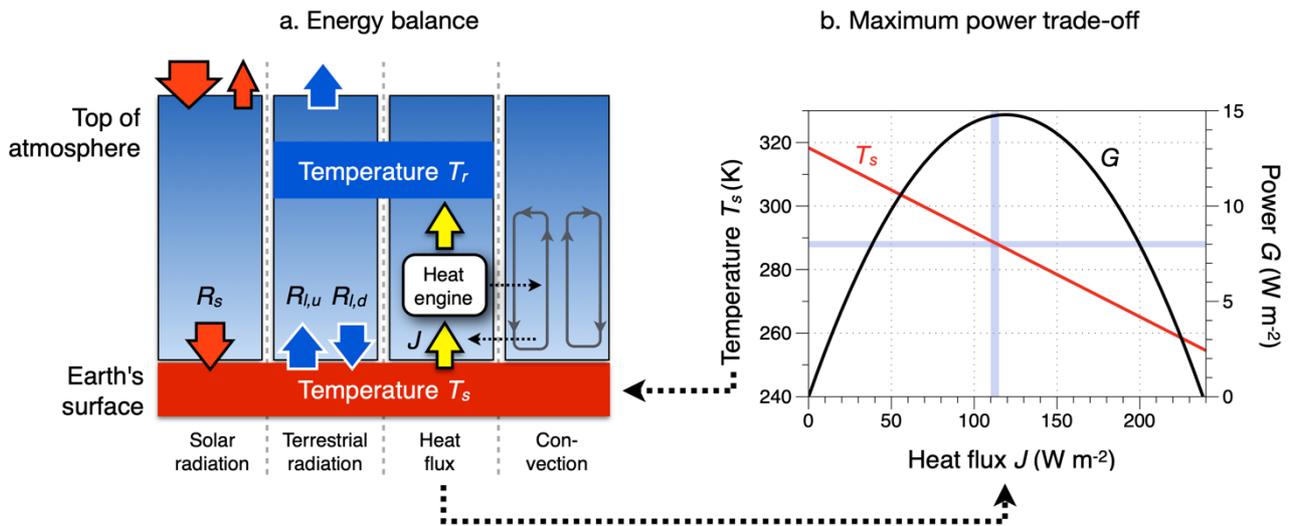

**FIGURE 3:** **(a.)** Schematic diagram illustrating the surface-atmosphere as a thermodynamic system with its associated energy fluxes. **(b.)** Illustration of the thermodynamic limit of maximum power that results when a heat engine situated in the atmosphere is linked with its consequences of moving air, providing a heat flux, cooling the surface, and thus lowering the efficiency of the heat engine.

I then describe a number of environmental datasets that are used in the evaluation. These datasets include the radiative forcing (including the absorbed solar radiation at the surface), a dataset on terrestrial photosynthesis, on evaporation, and on precipitation. These datasets are used to derive the low thermodynamic conversion efficiency from sunlight to carbohydrates during photosynthesis, they are used to show the tight linkage between photosynthesis and evaporation, and they are used to derive the maximum power limit and the predicted magnitude of evaporation. This predicted magnitude is then compared to an estimate from observations, and converted back to a conversion efficiency of photosynthesis. With this, the conversion efficiency of photosynthesis and its spatial variations can then be attributed to different physical factors of the environment.

*Maximum power limit on surface-atmosphere exchange*
The thermodynamic limit considered here deals with the vertical exchange of heat and mass between the surface and the atmosphere, which results from the heating of the surface by absorption of solar radiation. When solar radiation is absorbed and converted into heat, the overlying air is warmed, gains buoyancy, and rises. This exchanges air with characteristics representative of the conditions at the Earth's surface (in terms of its warmer temperature and its greater moisture and $CO_2$ content) with the colder and drier air of the atmosphere. It thus accomplishes the heat exchange between the surface and the atmosphere as well as the mass exchange of carbon dioxide and water. It represents the dominant mechanism for surface-atmosphere exchange, although it may be modulated by synoptic, atmospheric conditions, e.g., during periods of high wind speeds.

The magnitude of this surface-atmosphere exchange is limited by thermodynamics, much like a heat engine is limited by the Carnot limit (Figure 3). The convecting motion generated by the buoyancy of the warmed air near the surface is associated with kinetic energy, and the heat flux that generates this kinetic energy is the heat exchanged by buoyancy. Using the budgeting of energy fluxes that heat and cool the surface combined with the Carnot limit allows to derive a basic estimate on how the energy gained from the absorption of solar radiation and downwelling longwave radiation emitted by the atmosphere is partitioned into net radiative cooling and convective heat transport. In the following, I provide a brief summary of the approach, with details available in previously published work (Kleidon and Renner 2013a, Kleidon et al. 2014, Kleidon 2016, Kleidon and Renner 2017).

The power that is at best generated to accelerate air is described by the Carnot limit in the form of

$$G = J \cdot \frac{T_s - T_a}{T_s} \qquad (1)$$

where $G$ is the derived power (which generates convective, kinetic energy), $J$ is the heat flux associated with convection (in form of sensible and latent heat, $H$ and $LE$, with the former representing the transport of thermal energy and the latter being connected to the evaporation rate $E$ by the latent heat of vaporization, $L$), $T_s$ is the surface temperature, and $T_a$ is a representative temperature of the atmosphere. The temperature $T_a$ is taken to be the radiative temperature, $T_a = T_r$, which is determined from how much radiation is overall emitted to space, using the emission of a blackbody, $\sigma T_r^4$ (Note that in the global mean, the total emission is balanced by total absorption of solar radiation, although this can vary regionally due to heat transport). Thermodynamically, the radiative temperature is the coldest temperature by which the absorbed solar radiation can be emitted, so it represents a flux of maximum radiative entropy. The use of the radiative temperature in Eq. (1) is thus an upper bound to the power that can be generated. It is convenient to use as it is independent of the turbulent heat fluxes and how the absorbed solar radiation is partitioned at the surface. Note also that the expression of the Carnot limit in Eq. (1) can be derived directly from the first and second law in combination with the entropy budget (see, e.g., Appendix A in Kleidon and Renner 2013b, Kleidon 2016), so it is very general and does not require a particular assumption concerning the different steps of a specific thermodynamic cycle.

A critical aspect when applying the Carnot expression to the surface-atmosphere system is that the efficiency term in Eq. (1), $(T_s - T_a)/T_s$, is not independent of the heat flux $J$, but decreases the greater the turbulent heat flux is, because it cools the surface more vigorously. This interaction is captured by the surface energy balance. In a climatological mean state, it balances the heating by the absorption of solar radiation, $R_s$, and the downwelling terrestrial radiation, $R_{t,d}$, that was emitted by the atmosphere (what is commonly described as the atmospheric greenhouse effect) with the cooling by surface emission ($\sigma T_s^4$) and the turbulent heat fluxes $J$:

$$R_s + R_{t,d} = \sigma T_s^4 + J \qquad (2)$$

The terms on the left-hand side warm the surface, while the terms on the right-hand side cool the surface. Note that in the climatological mean state, heat storage changes can be neglected. When the net flux of terrestrial radiation, $R_{t,net} = \sigma T_s^4 - R_{t,d}$, is linearized in the form of $R_{t,net} = R_{t,0} + k_r (T_s - T_r)$, with $k_r$ being a linearization constant ($4 \sigma T_r^3$), the surface energy balance yields an expression that expresses the temperature difference $T_s - T_r$ in terms of the turbulent heat flux

$$T_s - T_r = \frac{R_s - R_{t,0} - J}{k_r} \qquad (3)$$

What we can see in this expression is how the temperature difference decreases with the greater heat flux $J$, so the efficiency term in the Carnot limit decreases correspondingly.

Hence, a maximum in power is obtained for a certain, optimum heat flux, $J_{opt}$, which is derived mathematically from $dG/dJ = 0$. Neglecting the dependency of $1/T_s$ on $J$ in the efficiency term yields a simple solution for this optimum heat flux of the form

$$J_{opt} = \frac{R_s - R_{t,0}}{2} \qquad (4)$$

This optimum heat flux can be further partitioned into its sensible and latent components assuming that evaporation is not limited by water availability. Then, the air at the surface can be assumed to stay saturated (i.e., in thermodynamic equilibrium with an open water surface), and the partitioning into the two fluxes is obtained by the so-called equilibrium partitioning (Schmidt 1915, Penman 1948, Priestley and Taylor 1972) given by

$$H_{opt} = \frac{\gamma}{s+\gamma} \cdot J_{opt} \quad (5)$$

and

$$LE_{opt} = \frac{s}{s+\gamma} \cdot J_{opt} \quad (6)$$

where $H_{opt}$ and $LE_{opt}$ are the optimum sensible and latent heat fluxes for the case where water does not limit evaporation, $\gamma$ is the so-called psychrometric constant ($\gamma$ = 65 Pa K$^{-1}$ for typical surface conditions), $s$ is the sensitivity of saturation vapor pressure to temperature, which depends strongly on temperature and can directly be expressed by the Clausius-Clapeyron relation, $s = de_{sat}/dT = L\, e_s/R_v\, T^2$, with $L$ is the latent heat of vaporization ($L \approx 2.5 \times 10^6$ J kg$^{-1}$ K$^{-1}$), $e_s$ the saturated vapor pressure at temperature $T$, and $R_v$ the gas constant for water vapor.

When water availability limits evaporation by precipitation input, $P$, so that $E \leq P$, the maximum power limits tells us that the sum of the heat fluxes $H + LE$ does not change. If we then assume that $E \approx P$, the partitioning changes to (as in Kleidon et al. 2014)

$$H = J_{opt} - LE \approx J_{opt} - LP \quad (7)$$

The thermodynamic limit of maximum power with respect to convective heat exchange from the surface to the atmosphere can thus be used to infer how the absorbed solar radiation is partitioned, it yields an associated surface temperature, as well as the magnitude of evaporation, which links to the associated exchange of mass at the surface. It requires the input of absorbed solar radiation, $R_s$, as well as a specification of the strength of the greenhouse effect to quantify $R_{t,0}$. With these two radiative fluxes, the turbulent heat flux $J$ is determined (Eq. 4), which is then further partitioned into sensible and latent heat, $H$ and $LE$ (Eqns. 4 and 5), from which evaporation is derived. Surface temperature $T_s$ can be estimated using Eq. 3, the expression of the optimum heat flux $J_{opt}$, and by deriving the radiative temperature $T_r$ from the radiative flux that is emitted to space.

*Environmental datasets*
A series of environmental datasets are used to determine the actual thermodynamic efficiency of photosynthesis on land, to relate it to evaporation, and to compare it to the thermodynamic limit. These datasets consist of the CERES radiative datasets (EBAF Version 4.1, Loeb et al., 2018;

Kato et al., 2018), the CASA-GFED dataset of gross carbon exchange (Version 3.0, Randerson et al. 2017; Ott 2020), that is, the net uptake of carbon dioxide by photosynthesis as simulated by the CASA model (Potter et al. 1993), the GPCP precipitation dataset (Adler et al. 2016), and the GLEAM evaporation dataset (Version 3, Miralles et al, 2011; Martens et al, 2017). The long-term climatological mean over the years 2003-2017 are used in the evaluation, a time period that is covered by all datasets. The datasets are summarized in Table 1, which also describes the symbols used to refer to the individual quantities. All datasets are based on satellite observations and are available as open access. The data was processed with the software tool "Climate Data Operators" (CDO, available at https://code.mpimet.mpg.de/projects/cdo) and plotted with python using Spyder Version 4.1.0 (available at https://www.spyder-ide.org). All datasets were brought to the same spatial resolution of 1° x 1° longitude/latitude using the CDO command "resample" for rescaling. This yields a total sample size of $n$ = 13274 grid points in the following evaluations.

*Conversions and Definitions*

In addition to using these datasets to quantify and test the maximum power limit, I analyze these datasets to quantify the thermodynamic efficiency of photosynthesis as well as the relationship to the evaporation rate.

| Dataset | Variables | DOI or Web Address | References |
| --- | --- | --- | --- |
| CASA-GFED (Figure 1a) | $A$: Gross ecosystem exchange (or, net photosynthesis), in gC m$^{-2}$ s$^{-1}$. | doi: 10.5067/VQPRALE26L20 | Randerson et al. (2017), Ott (2020) |
| CERES EBAF 4.1 (Figure 1b) | Radiative fluxes of solar and terrestrial radiation (in W m$^{-2}$): <br> $R_s$: Absorbed solar absorption at the surface; <br> $R_{t,d}$: Downwelling terrestrial radiation at the surface (greenhouse effect); <br> $R_{t,u}$: Surface emission (to infer surface temperature $T_s$, from $R_{t,u} = \sigma T_s^4$). | doi: 10.5067/Terra-Aqua/CERES/EBAF_L3B.004.1 <br><br> doi: 10.5067/TERRA-AQUA/CERES/EBAF-TOA_L3B004.1 | Loeb et al (2018), Kato et al. (2018) |
| GPCP (Figure 1c) | $P$: Precipitation, in mm month$^{-1}$. | doi:10.7289/V56971M6 | Adler et al. (2016) |
| GLEAM (Figure 1d) | Evaporative fluxes (in mm d$^{-1}$): <br> $E_t$: Transpiration by vegetative cover; <br> $E$: Total evaporative flux at the surface; <br> $E_{pot}$: Potential evaporation. | http://www.gleam.eu | Miralles et al. (2011), Martens et al. (2017) |

**Table 1: Overview of global datasets used to evaluate the thermodynamic conversion efficiency of terrestrial photosynthesis and its relation to radiative and water fluxes.**

First, I refer to the thermodynamic efficiency of photosynthesis, $\eta_{light}$, in units of %, as the ratio of generated chemical energy (in form of glucose and organic, reduced carbon) to absorbed solar radiation:

$$\eta_{light} = \frac{\Delta H_c \cdot A}{R_s} \qquad (8)$$

where $\Delta H_c A$ is the rate of photosynthesis in units of W m$^{-2}$. An average energy content of $\Delta H_c$ = 39 kJ gC$^{-1}$ is used to convert the photosynthetic rate $A$, which is typically given in units of grams of carbon per square meter of surface area and time, gC m$^{-2}$ s$^{-1}$ (see Table 1) into energy units. The value for $\Delta H_c$ is derived from the enthalpy of combustion of glucose of 2810 kJ mol$^{-1}$ and a carbon mol-weight of 72g of glucose (Atkins and de Paula, 2010).

Second, photosynthesis relates closely to evaporation rate, $E$, (see Fig. 1), a relationship that is well known and captured by the concept of the water use efficiency, $\varepsilon_{water} = A/E_t$ (where $E_t$ is transpiration, which in turn represents most of the evaporative flux on land), which is typically expressed in units of gC kgH$_2$O$^{-1}$. To make it consistent and comparable in terms of energy units, we use a slightly different way to calculate an efficiency $\eta_{water}$ (in units of %) here by setting the photosynthetic rate $\Delta H_c A$ (in energy units) in relation to the latent heat flux, $L E$,

$$\eta_{water} = \frac{\Delta H_c A}{LE} \qquad (9)$$

where $L$ is the latent heat of vaporization ($L \approx 2.5 \times 10^6$ J kgH$_2$O$^{-1}$).

Last, to evaluate photosynthetic efficiency in relation to geographic variations in climate, I use the aridity index of Budyko (1974; see also e.g., Milly 1994, and Roderick and Farquhar, 2011). The aridity index, $f_{arid}$, relates the demand for evaporation by the atmosphere in the absence of water limitation, a concept referred to as potential evaporation, $E_{pot}$, to the precipitation input, $P$:

$$f_{arid} = \frac{E_{pot}}{P} \qquad (10)$$

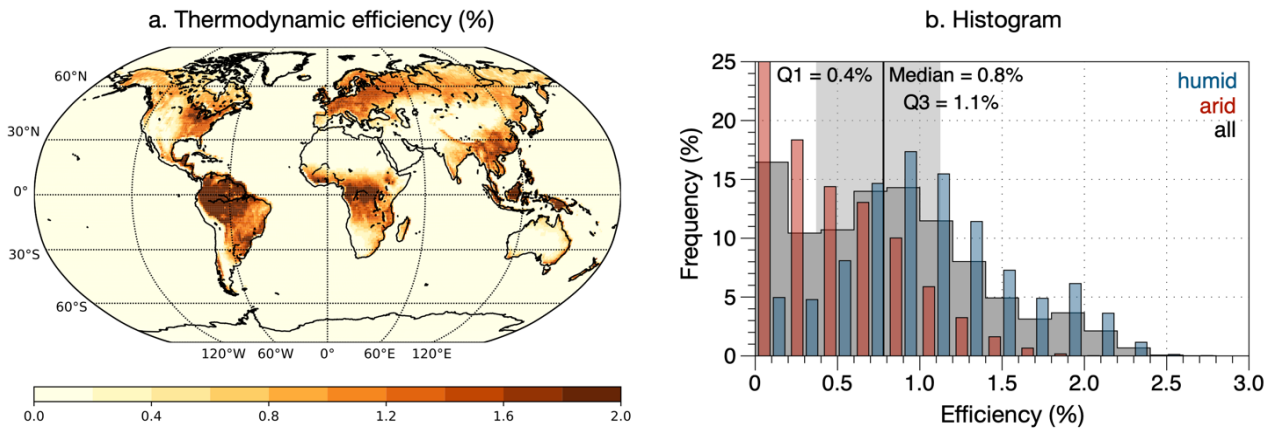

**FIGURE 4:** (a.) Map of thermodynamic efficiency in photosynthesis, $\eta_{light}$, calculated from the CASA-GFED and CERES datasets, as well as (b.) its frequency distribution in arid (red), humid (blue) and all (grey) regions. The quantiles of the distribution are marked in the histogram by the area shaded in light grey, with the quantiles provided in the figure.

An aridity index of $f_{arid} < 1$ represents a humid climate, as precipitation supplies more water than can be evaporated by the absorbed solar radiation. Arid regions are characterized by values of $f_{arid} > 1$, implying that more energy is absorbed in form of solar radiation than can be evaporated by the water input by precipitation. The aridity index is used in the analysis to differentiate between humid and arid regions in the analysis.

The processed datasets associated with the figures are available as an Excel spreadsheet as supplemental material.

**Results**

*Thermodynamic efficiency in terrestrial photosynthesis*

As a first step, the thermodynamic efficiency, $\eta_{light}$ (Eq. 8), was calculated by using the estimate for net photosynthesis (gross photosynthesis minus photorespiration, which is also referred to as gross primary productivity or the carbon fixation by photosynthesis) from the CASA-GFED dataset and the absorbed solar radiation from the CERES radiation dataset. The mean climatological variation is shown in Figure 4 as well as its frequency distribution of the different grid cells. The median efficiency of photosynthesis on land is 0.8%, with an interquartile range from 0.4% to 1.1%. One can see systematic geographic variations of the efficiency, with higher efficiencies in the humid regions such as tropical South America, Southeastern Asia, Eastern United States, and

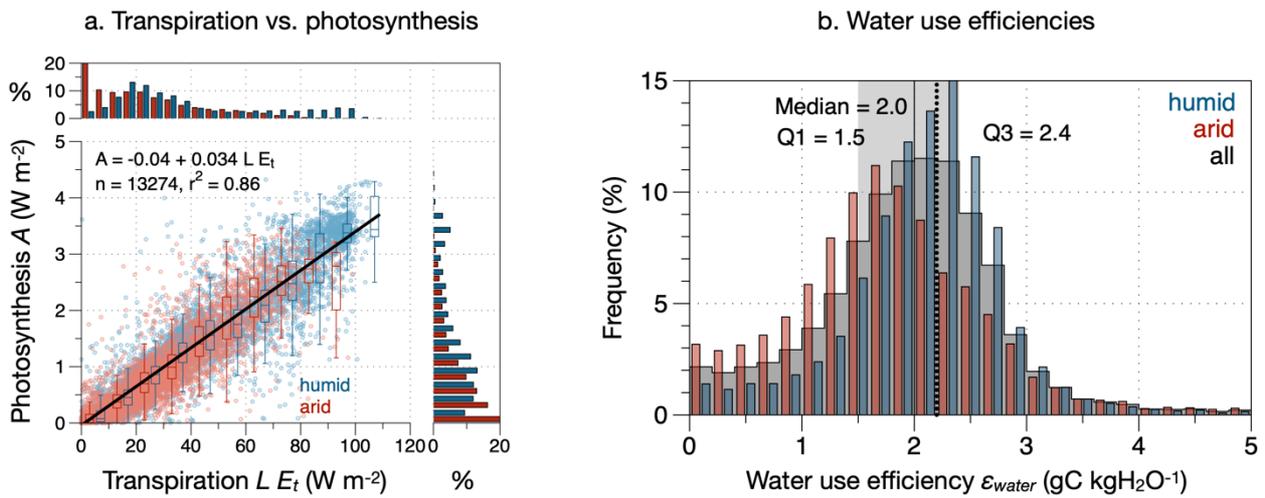

**FIGURE 5:** (a.) Scatterplot showing the close relationship of photosynthesis, *A*, (from the CASA-GFED dataset) and transpiration, $E_t$ (from the GLEAM dataset). Also shown at the top and at the right are the frequency distributions of the data points, with each data point representing one land grid cell of the maps shown in Fig. 1. (b.) Histograms showing the distribution of water use efficiencies, $\varepsilon_{water}$, derived directly from the ratio of $A/E_t$ for arid (red), humid (blue) and all (grey) regions. The dotted line in the histogram marks the value inferred from the linear regression, and the area shaded in grey marks the interquartile range.

Western Europe, and lower efficiencies in the semiarid and arid regions, such as sub-Saharan Africa, India, or the western United States. These low efficiencies are in agreement with those reported from earlier field studies (e.g., Monteith 1972, 1977).

Next, the photosynthesis dataset is compared to the transpiration rate that is contained in the GLEAM dataset (Figure 5). Note that transpiration, $E_t$, the contribution of plants to the total evaporative flux on land, $E$, represents about 88% in this dataset, estimated from the linear regression of these two fluxes, which yields a slope of 0.88 with a $r^2 = 0.87$. The photosynthetic rate in the CASA-GFED dataset correlates very strongly with the transpiration flux in the GLEAM dataset, with a linear regression yielding the best fit of $A = -0.04$ W m$^{-2}$ + 0.034 $L\,E_t$ ($r^2 = 0.86$). The tight correlation holds irrespective of whether the region is humid or arid, as indicated by the different colors in the scatter plot in Figure 5. Note that this correlation does not imply a causality in the sense that plant transpiration (or evaporation) drives carbon uptake. Rather it is argued here that this correlation reflects the simultaneous limitation of material exchange of carbon dioxide (*A*) and water ($E_t$).

In the regression, both fluxes were expressed as energy fluxes. The non-dimensional slope of 0.034 implies that for each joule of energy used in the evaporation of water, 0.034 joule of energy are fixed in carbohydrates. This corresponds to a water use efficiency of $\varepsilon_{water} = L/\Delta H_c \, \eta_{water} = 2.2$ gC kgH$_2$O$^{-1}$, implying that for each kilogram of evaporated water, 2.2 grams of carbon are fixed in carbohydrates. Expressed in terms of moles, this corresponds to a mean water loss of 276 mol for each mol of carbon fixed, which is consistent with the observed median of 317 mol H$_2$O/mol C (Q1 = 185 mol H$_2$O/mol C, Q3 = 545 mol H$_2$O/mol C, n = 87; Woods and Smith, 2010). When the water use efficiency is determined directly from the ratio of photosynthesis to transpiration (cf. Eq. 9 at each grid cell), it yields a distribution of values as shown in the histogram in Figure 5, with a median value of 2.0 gC kgH$_2$O$^{-1}$ and an interquantile range of 0.9 gC kgH$_2$O$^{-1}$. Note that the spread is also affected by the climatological conditions. In arid regions, the relative humidity is typically lower than in humid regions, which affects the gas exchange by increasing the water loss for the carbon gained. This effect is reflected in the shift of the histogram for arid regions towards lower values compared to humid regions in Fig. 5. These values correspond well to field observations across different ecosystems, which typically fall between 1 to 3 gC kgH$_2$O$^{-1}$ (Law et al. 2002, Tang et al. 2014).

In summary, this analysis of observational datasets confirms the low thermodynamic efficiency of photosynthesis and the well-established, tight connection to evaporation on land. In addition, the analysis provides information on the geographic variations and will serve as the basis for comparisons in the remaining part of the thermodynamic analysis.

*Maximum power and evaporation over land*

In the next part of the analysis, I quantify the thermodynamic limit of maximum power on surface-atmosphere exchange on land to infer the evaporation rate by using Eqns. 4, 6 and 7. As described above, the maximum power limit is a consequence of the tight linkage between the magnitude of the turbulent heat fluxes, $J$, with the surface temperature, $T_s$. In the following, I first illustrate this effect of the turbulent heat flux on surface temperatures before I then derive the evaporative flux and compare it to the GLEAM dataset.

To illustrate the extent to which the heat flux $J$ affects surface temperatures, I first computed $T_s$ in the absence of $J$ by setting $J = 0$ in Eqn 2 and then used this equation to calculate $T_s$. The temperature estimates are shown in Figure 6 (left) by the grey dots. The resulting temperatures correlate closely to those that are computed directly from the emitted radiation from the surface from the CERES dataset ($r^2 = 0.97$), but are generally much too warm (with a best fit of $y = -82 + 1.34 \, x$). When $J$ is determined from maximum power (red and blue dots in Fig. 6), these also correlate closely to those temperatures inferred from the CERES dataset ($r^2 = 0.97$), yet are on average about 16 K colder and agree much better with the observed surface temperatures (with a best fit of $y = -46 + 1.16\ x$). There is, however, a slight bias, with a cold bias in cold regions and a warm bias for warm regions. This bias is reflected in the regression slope of 1.16. Nevertheless, this evaluation of surface temperatures shows how important it is to account for the effect of the heat flux for surface temperatures and thus for the trade-off with the efficiency term in the Carnot limit (Eq. 1) in the maximization of power. Note that the radiative temperature that is used for $T_a$ in the limit is set by the total emitted radiation to space and is thus unaffected by the surface partitioning.

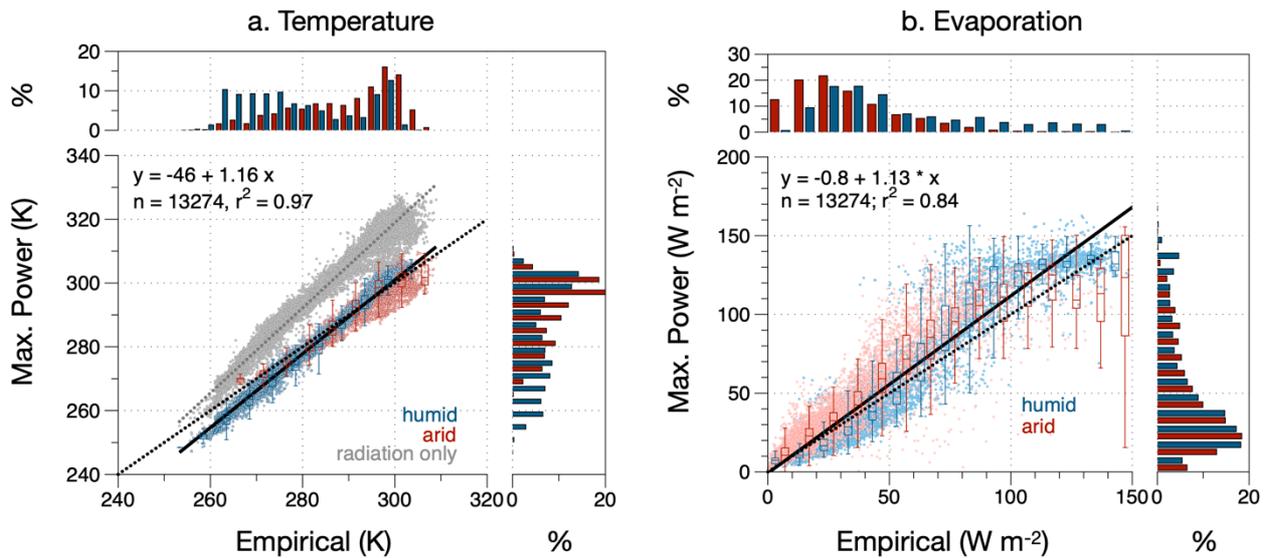

**FIGURE 6:** (a.) Estimated surface temperatures, $T_s$, and (b.) evaporation rates, $E$, from maximum power compared to the observation-based datasets. The grey dots in the left diagram show the estimate of surface temperatures when turbulent fluxes were absent ($J = 0$).

The evaporation rate derived from maximization of power is compared to the GLEAM dataset in Figure 6 (right). The two estimates correspond very closely to each other ($r^2 = 0.84$, with a best fit of $y = -0.8 + 1.13\ x$), although the maximum power estimate somewhat overestimates evaporation, as reflected in the regression slope of 1.13 being larger than one. The agreement is reasonably well for humid as well as for arid regions (blue and red dots in Figure 5, right). The overestimation to some effect is to be expected, as runoff is likely to be underestimated in this approach as the approximation $E \approx P$ in arid regions implies no runoff in arid regions. What these findings nevertheless support is that the maximum power limit appears to be a dominant constraint that sets the magnitude of the climatological evaporation patterns over land, confirming the outcome of an earlier study (Kleidon et al. 2014).

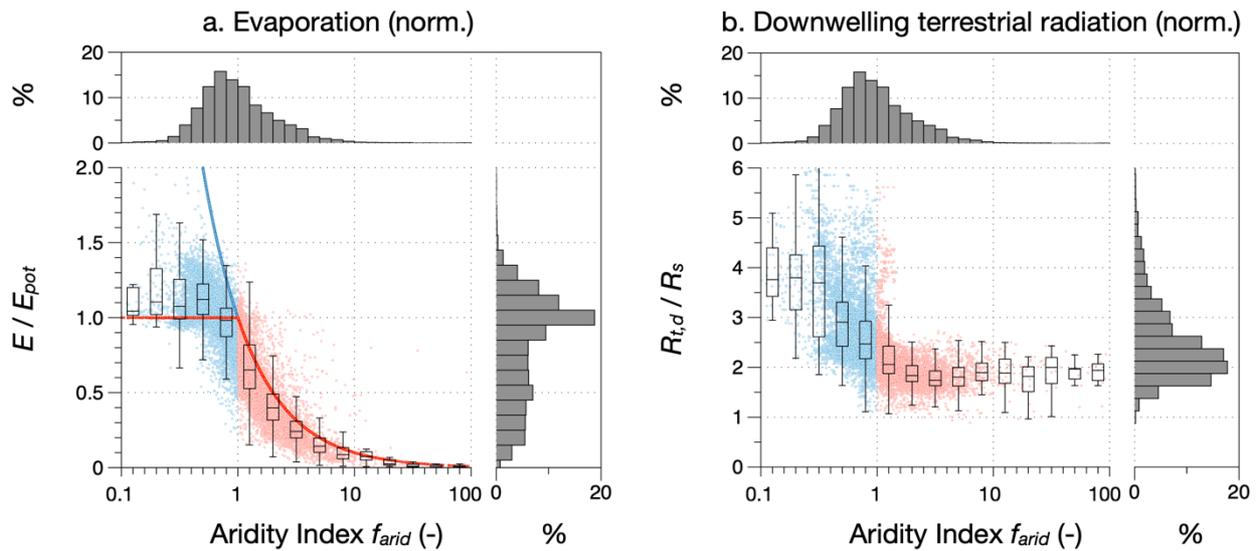

**FIGURE 7:** Geographic variation of (a.) evaporation, $E/E_{pot}$, from the GLEAM dataset, normalized by the potential evaporation rate, and (b.) downwelling terrestrial radiation, $R_{t,d}/R_s$, from the CERES dataset, normalized by the absorbed solar radiation, with the aridity index, $f_{arid}$. The red line in the left diagram shows the outcome from the maximum power estimate, and the blue line shows the corresponding value of precipitation, normalized by potential evaporation ($P/E_{pot}$).

*From evaporation to mass exchange limits and photosynthetic efficiency*

The analysis so far revealed findings that are largely consistent with previous research. So what do these insights now mean for what the dominant limitations are for photosynthesis?

I want to take one more step in the analysis before addressing this question. Since photosynthesis is so closely linked to evaporation by the gas exchange, let us first ask what the dominant limitation is that shapes evaporation rates, the mass exchange of water vapor between the surface and the atmosphere. This limitation has long been evaluated, and a suitable way to present this analysis is to use the aridity index, $f_{arid}$. This analysis is shown in Figure 7, which shows the normalized evaporation rate, $E/E_{pot}$, versus the aridity index, $f_{arid}$ (which is basically the inverse of a normalized precipitation rate, $P/E_{pot}$, see Eq. 10). Also shown in the figure are the corresponding estimates from the maximum power limit to show that this pattern is reproduced by this approach as well.

What Figure 7 shows is that for humid regions ($f_{arid} < 1$), evaporation takes place at its potential rate, which is set by the energy balance and not the water availability set by precipitation. This regime is typically referred to as the energy-limited regime of evaporation. In contrast, arid regions ($f_{arid} > 1$) represent the water-limited regime of the curve, reflected in normalized evaporation being close to the water supply by precipitation.  This classification between energy- and water-limited rates is also commonly applied to photosynthesis and ecosystem productivity, which is straightforward as photosynthesis and evaporation are coupled so strongly through gas exchange.  Yet, is it really energy that limits these rates?

In the so-called energy-limited regime, energy is far from being a limiting factor.  I show this in the right part of Figure 7, which shows the energy input by the downwelling, terrestrial radiation, $R_{t,d}$, to the surface, normalized by absorbed solar radiation, $R_s$.  What we can see is that $R_{t,d}$ generally supplies more energy to the surface energy balance than solar radiation does (the ratio of the two is > 1.5), and this is particularly the case in humid regions, where the ratio $R_{t,d}/R_s$ can be even larger.  This larger role of downwelling terrestrial radiation, reflecting the greenhouse effect, is well known, is needed to maintain habitable temperatures, and included in the comparison of temperatures in Figure 6.  Its amplification in humid regions is attributable to a greater concentration of water vapor as well as greater cloud cover, both representing the largest contributor to the greenhouse effect.

What this additional energy supply means is that the availability of radiative energy is not the limiting factor for the turbulent fluxes $J$.  It is rather the partitioning that is constrained, and it is constrained by the thermodynamic limit of maximum power that determines how much vertical exchange will take place.  If more of the energy supplied by the absorption of radiation would go into the heat flux $J$, this would lower the surface temperature further as well as the thermodynamic efficiency, lowering the generated power.  This power is, however, needed to drive the exchange of the heated, moistened, and $CO_2$ depleted air near the surface with the cooler, drier and $CO_2$ enriched air of the atmosphere.  The maximum power limit thus represents a thermodynamic constraint on mass exchange.  Consequently, evaporation in humid regions is not limited by energy - as commonly described - but rather by thermodynamics and material exchange.

To relate this thermodynamic material exchange constraint back to photosynthesis, I use the mean conversion factor of $\eta_{water}$ = 3.4% derived in the regression shown in Figure 5 and the 88% of the contribution by transpiration to the total evaporative flux and apply this to the evaporation estimate derived from maximum power.  This should be seen as a relatively rough quick-fix to get at photosynthetic rates and would need further investigations whether this conversion rate of water vapor exchange to $CO_2$ exchange can be explained more physically, or whether it is dominated by physiology.  The resulting geographic distribution of estimated photosynthetic rates is shown in Figure 8, which also shows the comparison of derived thermodynamic efficiencies to those derived from the CASA-GFED dataset. Since photosynthesis and evaporation correlate so strongly, it is not surprising to see that this conversion back into a photosynthetic rate can reproduce the CASA-GFED dataset quite well ($r^2 = 0.66$, best fit of $y = 0.38 + 0.89\,x$), with a systematic bias to overestimate low photosynthetic rates. The agreement is stronger in water-limited, arid regions ($r^2 = 0.73$, $n = 5676$, best fit $y = 0.37 + 0.98\,x$) than in humid regions ($r^2 = 0.56$, $n = 7598$, best fit $y = 0.37 + 0.88\,x$).

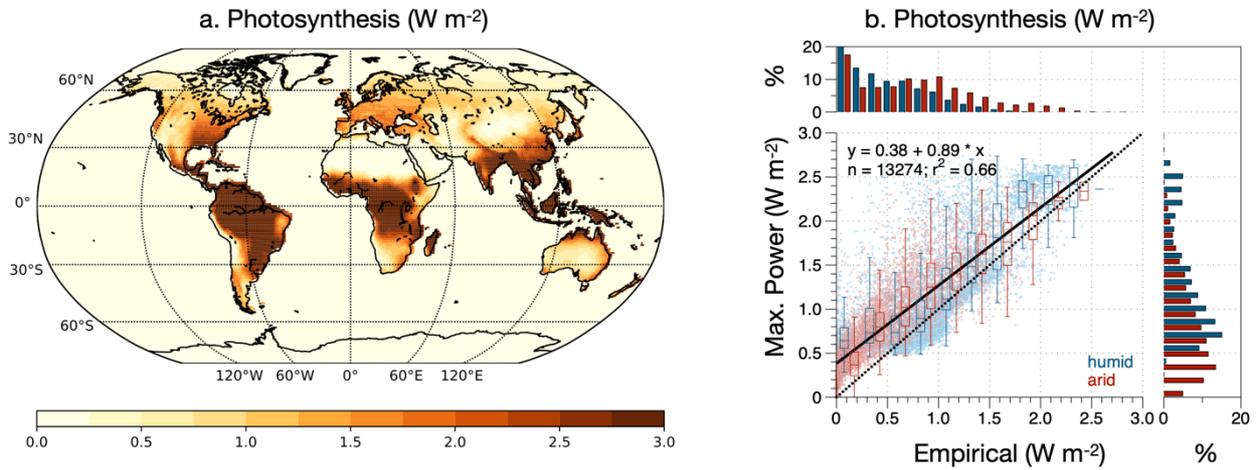

**FIGURE 8**: (a.) Photosynthetic rate, $A$, inferred from the evaporation rate, $E_t$, derived from the maximum power limit, using a constant water use efficiency of 3.4%, as derived from the linear regression shown in Figure 5. (b.) Scatterplot showing the grid-based comparison of the photosynthetic rates to those inferred from the CASA-GFED dataset.

We can next use this rather rough estimate of photosynthesis to better understand why its thermodynamic efficiency $\eta_{light}$ is so low and what the major factors are that determine its geographic variations. The major factors that determine the variations in thermodynamic efficiencies derived here from maximum power are the same as the ones that determine the variations in evaporation rate, as shown in Figure 7. To demonstrate this, the geographic variation of thermodynamic efficiency is related in Figure 9 to the prefactor, $s/(s + \gamma)$, used in the partitioning of the optimum turbulent heat flux into the evaporation rate (Eq. 6) for humid regions and to precipitation, $P/E_{pot}$ (which is equivalent to the inverse of the aridity index, $f_{arid}$, Eq. 10), normalized by the potential evaporation rate, for arid regions.

These two correlates shown in Figure 9 support the notion that in humid regions, the material exchange needed for photosynthesis is thermodynamically constrained while in arid regions, it is the availability of water by precipitation input that constrains the exchange. This view can explain the low thermodynamic efficiency of photosynthesis in natural environments and its geographic variation fairly well. The low thermodynamic efficiency thus seems not to be caused by inefficient energy conversions, but inefficient material exchange.

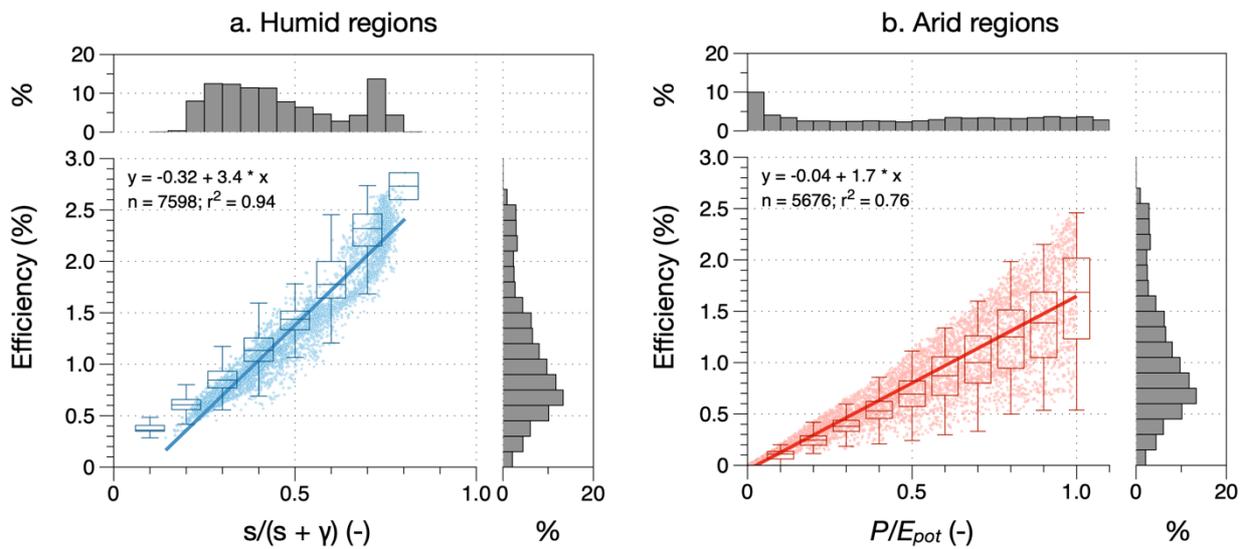

**FIGURE 9: Photosynthetic efficiency, $\eta_{light}$, inferred from the maximum power as a function of (a.) the equilibrium partitioning factor $s/(s + \gamma)$ for humid regions and (b.) normalized precipitation, $P/E_{pot}$, which is equal to the inverse of the aridity index $f_{arid}$.**

**Discussion**

The analysis performed here is, of course, held at a highly simplified level, leaving out many factors that could affect the results to some extent. For instance, annual means were used in the analysis, yet some of the variables used to estimate the flux partitioning and evaporation rates vary substantially at diurnal and seasonal scales, and the associated covariances may not average out in the climatological mean. This aspect could be refined in a more detailed analysis. Such an analysis would be necessarily more complex, as it would, for instance, require a more detailed representation of the soil water balance to infer water limitation, rather than the simple separation between arid and humid regions and the use of potential evaporation rate to precipitation. Also, the conversion of the evaporation rates determined from maximum power back to photosynthetic rates were performed with an ad-hoc, empirical water use efficiency, without a physical basis that would determine its value. In principle, this value should also be possible to derive from the maximum power limit, but it would require a more detailed representation of the exchange process of water vapor and carbon dioxide between the interior of the leaves and the surrounding atmosphere, along with the temperature gradients that drive the buoyant exchanges. These aspects, among others, would need further work in the future to improve the analysis performed here and substantiate the interpretation.

The analysis performed here reproduces the low thermodynamic efficiency of photosynthesis and the relationship between terrestrial photosynthesis and evaporation. By linking these two, well-known aspects with the maximum power limit on land-atmosphere exchange, it offers a novel interpretation of the low thermodynamic efficiency of photosynthesis. Instead of viewing the energy conversion process of photosynthesis as being thermodynamically constrained, this novel interpretation suggests that it is the mass exchange of carbon dioxide between vegetation and the atmosphere that is thermodynamically limited. As this mass exchange is driven by the heating due to absorption of solar radiation during the day, this results in a correlation between photosynthesis and solar radiation in humid regions that looks as if the photosynthetic rate was light limited. However, thermodynamics tells us that photosynthesis could be a lot more efficient in converting light, so that the availability of light is not the constraining factor. This led to the proposition formulated here that photosynthetic rates in natural ecosystems are not directly limited by sunlight, but rather by the transport of carbon dioxide between the canopy and the atmosphere. It thus forms an indirect, thermodynamic limitation that looks like as if it were a direct limitation by light availability. This latter point was substantiated here by invoking the thermodynamic limit of maximum power, which sets an upper limit to the intensity of convective exchange between the surface and the atmosphere, and that was used here to infer evaporation rates from absorbed solar radiation. Given that the gas exchange of water vapor and carbon dioxide are closely tied, which was represented here by using a fixed, empirical water use efficiency to convert the evaporation rate from maximum power to a carbon uptake rate, this can then explain the observed, very low thermodynamic conversion efficiency of photosynthesis. This low efficiency results not from an inefficient use of light, but because the physical transport to the photosynthesizing tissues in plant canopies appears to be inefficient in transporting more reactants. Yet, as the evaporation rate here is predicted from the thermodynamic limit, it implies that this is as much material exchange as can be generated from the solar radiative heating of the land surface.

This line of reasoning leads to the interpretation that the highly productive ecosystems of the humid tropics are primarily limited by physical transport, which may be predicted by purely physical variables, specifically solar radiation in conjunction with the maximum power limit. This interpretation is consistent with the long-standing and well-established notion that vegetation types and their productivity strongly correlate with physical, climatological variables irrespective of their species composition and evolutionary histories, resulting in biogeographical patterns and characteristic climate types. It is also consistent with observations (and theory) that elevated levels of carbon dioxide in the atmosphere result in greater photosynthetic rates, a well-established effect referred to as the $CO_2$ fertilization effect, because a higher atmospheric concentration of carbon dioxide would cause a greater exchange of $CO_2$ with the vegetated surface. If light was the limiting factor at the ecosystem scale (in contrast to the scale of chloroplasts, for which it is well established that the process is colimited by light (electron transport) and $CO_2$, Farquhar et al. 1980), photosynthesis would not be expected to respond to atmospheric levels of $CO_2$. The interpretation provided here, however, appears notably different than previous interpretations of the low photosynthetic efficiency in the agricultural literature (Long et al. 2006, Zhu et al. 2010). In these studies, the low efficiency of photosynthesis was attributed to photorespiratory loss. This explanation may, in fact, not contradict the explanation given here, but merely represent the way by which plants cope with the ATP harvested by the chloroplasts given that the material exchange of $CO_2$ limits the incorporation of the acquired energy into carbohydrates. This speculation, however, would require further evaluation.

So it would seem that the interpretation proposed here, that thermodynamics limits biotic activity through its constraint on mass exchange with the environment, is at least plausible. Further evaluations would be needed for this hypothesis to be substantiated. This could, for instance, involve on the theoretical side an extension of the thermodynamic approach to predict the slope of the correlation between carbon and water fluxes (which links to the so-called aerodynamic conductance that describes surface-atmosphere exchange), an aspect for which the empirically-found conversion factor of $\eta_{water}$ was used here. When the predictions are to be further evaluated with observations, an important aspect to consider seems to be the covariation of environmental processes, which are likely mirrored in how plants adapted their ecophysiological functioning to the environment. Furthermore, the material exchange limitation described here applies to ecosystems in the natural environment at the larger scale, so the insights may not apply in the same way when evaluating individual plants, which take up much less carbon when compared to whole ecosystems. It would thus seem that future evaluations should focus on the larger ecosystem scale, account for the covariations among variables, and focus specifically at the exchange across the biosphere-atmosphere interface. The use of eddy covariance measurements may help in doing so as they measure surface-atmosphere exchange for more than an individual plant and include the natural covariations among processes.

If we accept this proposition that the major constraint on biotic activity on Earth is related to the thermodynamic constraints on material exchange, what would be the broader implications? What I want to do in the following is to speculate on such broader implications, assuming that mass exchange is indeed the limiting factor. What this would mean is that when we aim to understand the constraints of living systems and how, and how much, these convert energy to stay metabolically active, the thermodynamic context of the environment plays the central role. It demands to take a systems' perspective, in which living organisms are embedded in their thermodynamic environment. While the dissipative activity of an individual living organism may depend on a variety of biological factors, the perspective described here suggests that the activity of the whole ecosystem may very well be thermodynamically constrained, essentially irrespective of the individuals that form the ecosystem. The thermodynamic constraint, however, does not materialize directly through the energy conversion processes within the individuals, but rather indirectly through the ability of the environment to supply the food and remove the waste products, to use the terminology of Schrödinger's (1944) seminal work on what is life. This, in turn, would be consistent with scaling laws associated with individual trees in forest ecosystems that can be derived from a resource constraint (Enquist et al., 1999). It is consistent with the interpretation here in that it is the environment that sets the level of biotic activity of the ecosystem through its magnitude of mass exchange, rather than the specific behavior of individuals and their abundances. Thus, thermodynamics can constrain biotic activity, not of the individual, but rather of the whole biosphere.

We can next ask how relevant such thermodynamic constraints are for biological systems more generally. In the above, we used the thermodynamic limit of maximum power to infer the extent of convective mass exchange between the surface and the atmosphere. The ability of this approach to reproduce observation-based estimates of evaporation, as reflected in the close agreement with observation-based datasets in Figure 6, suggests that convective mass exchange evolves to and operates near this thermodynamic limit. If we then transfer this emergent behavior of convective motion to biologic activity as a dissipative process, it would, likewise, suggest that the activity of an ecosystem would evolve to, and eventually reach a thermodynamic limit of maximum activity, translating into maximized photosynthetic rates. These rates would then be predictable from the thermodynamic limit, similar to how rates of terrestrial photosynthesis were inferred in the above analysis. This notion is consistent to what Lotka already formulated in his works about a hundred years ago (Lotka 1922a,b), namely, that evolution by natural selection should favor these organism of greater power, eventually resulting in biologic systems maximizing their power (with power being the generation rate of chemical free energy). The evolution to this state of maximum productivity would, however, be associated with a much longer time scale than that of atmospheric motion, so it would represent more of an evolutionary direction rather than a fast, emergent outcome.

A critical role of mass exchange as a constraining factor for life would also have implications for how life may have emerged and which planetary environments are habitable. One theory of the origin of life proposes that life emerged at hydrothermal vents at the ocean floor (Wächtershäuser 1990). Without getting into the details of the chemical conditions at such sites, this proposition is consistent with the notion that mass exchange plays a critical role. The difference of these sites compared to land is that at hydrothermal vents, the mass exchange is driven by the comparably high heat flux at the plate boundaries from the interior. The strength of this mass exchange at hydrothermal vents also appears to be thermodynamically constrained (Jupp and Schultz, 2000). It would thus seem that the habitability of a planet relates to its ability to sustain mass exchange of reactants and products, hence showing a substantial level of dissipative activity. Typically, planetary habitability is simply related to the temperature at which liquid water can be maintained (e.g., Seager 2013). What this mass exchange constraint to life would suggest that additionally, the planetary environment would need to be sufficiently thermodynamically active to allow dissipative processes such as mass exchange to take place. On Earth, this condition is met by temperatures that allow liquid and gaseous phases of water to coexist, and its distance to the Sun is such that the heating by solar absorption is sufficient to maintain substantial levels of mass exchange.

As a last step, we can make a link between biotic activity and the consequences it has for the planetary environment in thermodynamic terms. To provide a template of what to expect, I first want to go back to the example of atmospheric convection. In this case, the effect of maximum power is an acceleration of the second law of thermodynamics at the planetary scale. By redistributing the emission of the absorbed solar radiation from the warmer surface to the colder temperatures of the atmosphere, it results in higher entropy of the emitted radiation, permitting greater entropy production and dissipative activity within the planetary system. In other words, the presence of atmospheric convection has an impact on the thermodynamic behavior of the whole planet, allowing for it to become more dissipative than in its absence. The process of convection would occur irrespective of the presence or absence of life. The heat fluxes inferred from the maximum power limit can, for instance, reproduce the observed heat fluxes over tropical rainforest as well as over adjacent, bare ground of agricultural fields in Southeastern Amazonia (Conte et al., 2019). Yet, it demonstrates how the inclusion of a dissipative process (convection) evolves to a thermodynamic limit, and how this enhances the thermodynamic activity of the whole planet.

When we apply this template to biotic activity as a dissipative process, would we expect the similar thermodynamic outcome as for convection? That is, is Earth with a biosphere more thermodynamically active than a lifeless planet? First, when the biosphere maximizes its productivity, it would also maximize the dissipation of the process that we call life. But would it make the Earth more thermodynamically more active?

The link to the planetary environment is established through the biotic effects on the environment. There are several kinds of such effects. Rainforests, for instance, absorb more solar radiation than bare ground because their canopies absorb more sunlight. This enhanced absorption would drive more convective exchange. They evaporate more water back into the atmosphere by being able to reach water stored deeper within the soil through their root systems. This results in higher rates of evaporation during periods when water can limit evaporation, such as dry seasons, affecting the partitioning of the absorbed solar radiation into evaporation and the sensible heat flux (Eqns. 5 – 7). The magnitude of convective exchange, as predicted by the maximum power limit, is nevertheless unaffected (Conte et al. 2019). By taking up and storing carbon dioxide, the biosphere actively performs the chemical work to draw down the concentration of this atmospheric greenhouse gas. Photosynthesis generates an order of magnitude more chemical energy than all other geochemical processes on Earth (Kleidon, 2016), so it is mostly through life that the chemical composition of the atmosphere, and hence its radiative properties, has been transformed throughout Earth's history. This, in turn, affects the radiative environment in which convection occurs. Hence, biologic activity impacts the physical constraints it is subjected to, particularly regarding the radiative forcing. We could thus imagine that by maximizing its activity, the biosphere would then be able to maintain a state in which the planetary environment is regulated to a state most conducive to biotic activity. Such behavior is then very close to what was postulated by the Gaia hypothesis by Lovelock and Margulis (1974), which states that the Earth's environment is regulated by life for its own benefit.

While these broader implications are certainly speculative and would need much more detailed evaluation, thermodynamics and its limits, jointly with an Earth system perspective in which mass is being converted, exchanged, and transported, is needed to provide the essential foundation that would make these implications plausible and is needed to evaluate these. It would be through the effects and interactions of biological systems with their environment from which maximum power states emerge, and it is the thermodynamic limit on physical mass exchange that would limit their activity and make these highly complex systems relatively simple and predictable in terms of their overall thermodynamic activity.

**Summary and Conclusions**

In this paper, I evaluated the question about the factors that limit photosynthesis and whether thermodynamics plays a role in setting these limits. This question is motivated by the long standing observation that the photosynthetic rates of terrestrial ecosystems is much lower than the limits inferred from thermodynamics. What I proposed here is that thermodynamics acts indirectly as a constraint to photosynthesis. This constraint is indirect because it does not limit the energy conversion process from light to carbohydrates directly, but rather the gas exchange of water vapor and carbon dioxide between the plant canopies and the atmosphere. I used global data sets on photosynthesis, radiation, and hydrologic fluxes to evaluate and support this interpretation quantitatively. The large-scale geographic variations in terrestrial photosynthesis is fairly well explained by this approach, although it leaves the question open as to which extent the magnitude of the water use efficiency is set by the physical exchange process as well. Yet, this interpretation can explain why the geographic patterns of photosynthesis are highly predictable from climatological variables, as these variables are associated with the thermodynamic limit of surface-atmosphere exchange.

At the broader level, this interpretation emphasizes that life does not just need an energy source to stay active, it also needs a thermodynamically active environment that exchanges the reactants and products. This exchange involves the energy conversion from heat to motion, and is also subject to thermodynamic conversion constraints. Hence, to understand and interpret constraints on the activity of life, its origin and evolution, it would seem that it requires a thermodynamic Earth system perspective.

**Acknowledgements**

I thank two reviewers for their constructive reviews that helped to improve the manuscript and the scientists of the CASA-GFED, CERES-EBAF, GPCP and GLEAM datasets for making their datasets openly accessible.

**Supplemental Material**

A text file with the data points to generate the plots and maps is available as well as the processing scripts at https://dx.doi.org/10.17617/3.4q.